\documentclass[aps,pre,epsfig,floats,twocolumn,superscriptaddress,amssymb,amsmath,showpacs,article]{revtex4}
\usepackage{graphicx}
\usepackage{isorot}
\usepackage{color}
\begin{document}
\title{Microtubule Dynamics and Oscillating State for Mitotic Spindle}

\author{Safura Rashid-Shomali}
\email{rshomali@iasbs.ac.ir}
\affiliation{Institute for Advanced Studies in Basic Sciences,
Zanjan 45195-1159, Iran}

\author{Ali Najafi}
\email{najafi@znu.ac.ir}
\affiliation{Zanjan University, Zanjan 313, Iran}

\date{\today}

\begin{abstract}
We present a physical mechanism that can cause the mitotic spindle to oscillate.
The driving force for this mechanism emerges from the polymerization of astral
microtubules interacting with the cell cortex. We show that Brownian ratchet model for
growing microtubules reaching the cell cortex, mediate an effective mass to the spindle body and therefore force it to oscillate.
We compare the predictions of this mechanism with the previous mechanisms which were based on the effects of motor proteins. Finally
we combine the effects of microtubules polymerization and motor proteins, and present the detailed phase diagram for possible oscillating states.
\end{abstract}
\maketitle
\section{Introduction}
Establishment of geometrical polarity during mitosis which is a result of asymmetric cell division is a fundamental fact in many live systems. The understanding of underlying molecular mechanisms for this phenomena is an important and challenging issue in biophysics \cite{cellintro}.
Including cell division, various functionalities of cell are related to the microtubules, the most rigid filaments of eukaryotic cytoskeleton. These
filaments are responsible for the mechanical properties of cells \cite{cell,MTint1}. The functional properties of microtubules are determined by the structures of their monomers and the interaction
mechanism packing them into rigid polymers. These long and hollow cylinders are made from $\alpha-\beta$ tubulins, proteins in the eukaryotic cells~\cite{howardbook}.

Microtubules growing from two organizing centers inside the cell, form a reliable scaffold for the mitotic spindle and make the mechanics of the cell.
In the animal cells during mitosis, microtubules are nucleated from centrosomes, organizing centers near nucleus \cite{cell}. Some microtubules grow from
one centrosome toward another and construct a firm bridge between two
centrosomes. These microtubules with two centrosomes, form the
complicated structure of the mitotic spindle which plays an
important role in cell division. Microtubules also, are required for
divorcing the chromosomes at the end of mitosis~\cite{chromosome
oscillation}. Some other microtubules, astral microtubules, growing from centrosomes, reach the cell cortex and push the spindle body. They are absolutely required for correct positioning of mitotic spindle~\cite{tolic-MTpusshing}.

It is understood that the spindle motion during the cell division is responsible for asymmetric cell division.
To gain insight into the cellular mechanisms by which the spindle oscillates and repositiones during the mitosis, extensive studies are carried out on the \emph{C. Elegans} embryo \cite{grillexper}. All experiments have clarified the importance of microtubules and their polymerization.

Considering a solid spindle body and astral microtubules, interacting with cell cortex, two different deriving mechanisms for spindle
motion are possible. The first is the pushing and pulling forces related to the motor proteins connecting the microtubule tips to the
cell cortex and the second is the pushing mechanism due to the microtubule polymerization. The first mechanism is studied by S. W. Grill {\it et al.} \cite{julicher}, where, they have shown that the resulting forces make the spindle oscillation possible. Regarding the second mechanism, the existence of pushing forces is experimentally investigated and theoretically studied \cite{MTpushing,MTkozlow, PNAS_dogterom}. In such systems the effects of
microtubule bending and buckling are discussed \cite{nedelec2,howard,Science_dogterom}.\newline

In this article we concentrate on spindle body motion with deriving forces originated from the microtubule and cell cortex
interaction. We present a simplified theoretical model with minimum requirement to generate an stable oscillating state for spindle body.
In section II, we briefly review the governing equations for the dynamical instability of microtubules and also
Brownian ratchet model\cite{brownianratchet,prostBrownianRatchet}. In section III, we show that the interaction of microtubules with cell cortex can produce an overall effective mass for the spindle body and force it to oscillate. The effect of the
molecular motors is considered in section IV. Discussion about our results and concluding remarks
are presented in section V.

\section{Microtubule's dynamics and Brownian ratchet mechanism}
Let us consider a bundle of stiff and noninteracting microtubules growing from a centrosome (nucleating center).
The life of these microtubules can be seen as two different phases: the
growth and the shrinkage phases.
To describe the stochastic transition of microtubules between two phases in a simplified model, four parameters are defined.
Two transition rates, $f_{cat}$ and $f_{res}$, the catastrophe and rescue frequencies that quantify the rates by which the microtubules change their growing and shrinking phases and also average growth and shrinkage velocities for microtubules at the bulk which are shown by $v_{g}$ and $v_{s}$ respectively.
Denoting by $n^{+}(x,t)$, the number density of growing microtubules at position $x$ and by $n^{-}(x,t)$, the number density of shrinking microtubules, the dynamical equations can be written in the following form \cite{dogterom1}:
\begin{eqnarray}
\frac{d}{dt}n^{+}(x,t)&=&\frac{v_{g}}{l}\delta n^{+}(x,t)-f_{cat}n^{+}(x,t)+f_{res}n^{-}(x,t)\nonumber,\\
\frac{d}{dt}n^{-}(x,t)&=&\frac{v_{s}}{l}\delta n^{-}(x,t)-f_{res}n^{-}(x,t)+f_{cat}n^{+}(x,t)\nonumber,
\end{eqnarray}
where $\delta n^{\pm}=n^{\pm}(x\mp l,t)-n^{\pm}(x,t)$ and $l$ stands for the monomer length.

Despite the above equations for microtubule dynamics in the bulk, the microtubules have
different evolving equations on the centrosome (nucleation cite), and also on the cell cortex. To express the
dynamical equations for the microtubules on the centrosome at $x=0$, we define $\nu$ as nucleation rate . The nucleation rate is a rate
by which free inactive sites on the centrosome change to active sites and grow by absorbing tubulins. Now the boundary equations at
the centrosome read \cite{dogterom2}:
\begin{eqnarray}\label{nucleation}
\frac{d}{dt}n^{+}(0,t)&=&\nu n^{-}(0,t)-\frac{v_{g}}{l}n^{+}(0,t),\nonumber\\
\frac{d}{dt}n^{-}(0,t)&=&-\nu n^{-}(0,t)+\frac{v_{s}}{l}n^{-}(l,t),
\end{eqnarray}
Correspondingly the boundary equations at the position of the cell cortex that is located at a fixed position $x=L$, can be written as\cite{dogterom2}:
\begin{eqnarray}
\frac{d}{dt}n^{+}(L,t)&=&\frac{v_{g}}{l}n^{+}(L-l,t)-
f_{wall}n^{+}(L,t),\nonumber\\
\frac{d}{dt}n^{-}(L,t)&=&-\frac{v_{s}}{l}n^{-}(L,t)+
f_{wall}n^{+}(L,t),
\label{wallboundary}
\end{eqnarray}
where $f_{wall}$ represents the catastrophe frequency on the obstacle.

For the next step, we present the steady
 state solutions to the above system of equations. Using matrix techniques for difference equations and for a wall that is fixed  at
position $L=L_0$, the solutions read \cite{dogterom2}:
\begin{eqnarray}
n^{+}(x)&=&A\exp(
-\frac{x}{\lambda})~~~~~~~~~~~~0\leq x < L_{0}\nonumber\\
n^{-}(x)&=&\frac{v_{g}}{v_{s}}A\exp(
-\frac{x}{\lambda})~~~~~~~~0 < x\leq L_{0},
\end{eqnarray}
and the boundary values are:
\begin{equation}
n^{-}(0)=\frac{v_{g}}{\nu } n^{+}(0),~~~n^{+}(L_{0})=\frac{v_{s}}{f_{wall} }n^{-}(L_{0}),
\label{equilibriumvalue}
\end{equation}
with
\begin{equation}
A=\left(\lambda(1+\frac{v_{g}}{v_{s}})(1-e^{
-\frac{L_{0}}{\lambda}})+\frac{v_{g}}{\nu}\left(1+\frac{\nu}{f_{wall}}e^{
-\frac{L_{0}}{\lambda}}\right)\right)^{-1},
\end{equation}
where $\lambda=\left(\frac{f_{cat}}{v_{g}}-\frac{f_{res}}{v_{s}}\right)^{-1}$.

The growth and shrinkage velocities that we have used in above descriptions, are assumed to be the velocities for free microtubules
in the bulk. For a microtubule tip reaching an obstacle, polymerization process can persist by pushing which repositions centrosome (nucleation cite). Brownian ratchet model is a physical mechanism that describes the dynamics of this process \cite{prostBrownianRatchet,brownianratchet,tanase}.
Denoting by $F$, the force that the wall exerts on the tip of a microtubule, the following linear response, force velocity relation can be written:
\begin{equation}
F=F_{stall}\left(1-\frac{v}{v_g}\right),
\end{equation}
where $v$ is the polymerization velocity in the presence of the load and $v_g$ stands for the polymerization velocity in the absence of the load.
The stall force $F_{stall}$, is the threshold force that stops the polymerization process. The stall force is given by $F_{stall}=\frac{k_BT}{l}\left(1-\frac{k_{off}^{0}}{k_{on}^{0}}\right)$, where $k_{on}^{0}$ and
$k_{off}^{0}$ are polymerization and depolymerization rates. One should note that the bulk polymerization velocity is related to the microscopic rates by:
\begin{equation}
v_{g}=l\left(k_{on}^{0}-k_{off}^{0}\right).
\end{equation}
Defining friction coefficient by $\xi_0=\frac{F_{stall}}{v_g}$, we can rewrite the force velocity as: $F=\xi_0\left(v_g-v\right)$.

In the next section we will use above description about microtubule dynamics and Brownian ratchet mechanism to analyze the dynamics
of spindle body.

\section{Dynamics of spindle body}
Astral microtubules with the mitotic spindle constitute our one
dimensional model shown in Fig.~\ref{fig:shape1}. In our model, spindle structure, behaves like a rigid body. Dynamical instability of astral microtubules makes stochastic transition between growth and shrinkage phases.
During the polymerization process, each microtubule that reaches the wall, pushes it with force $F$ that obey the Brownian
ratchet response function \cite{brownianratchet}. Additionally we assume that the cortex is a fixed
and rigid wall that does not displace or deform
during spindle motion. The mitotic spindle then, feels the reaction of microtubule
forces on both sides. The reaction force from the left (right) hand side wall is proportional to the
number of microtubules which have reached the left (right) wall. Denoting the number of attached microtubules in the left side by
$N_l$ and the corresponding number at right side by $N_r$, and using the linearized Brownian ratchet response function,
the net polymerization force acting on the spindle body can be written as:
\begin{figure}
\includegraphics[width=.8\columnwidth]{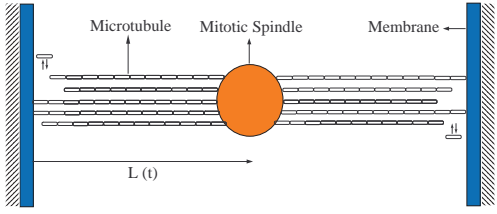}
\caption{(color online). One dimensional modeling of spindle body and astral microtubules in mitosis; two
centrosomes and the microtubules between them assumed to be rigid
and form spindle. The cell cortex is fixed and the net force due to the pushing of attached microtubules cause the spindle to move.}
\label{fig:shape1}
\end{figure}

\begin{equation}\label{Fpol}
F_{pol}=\left(N_{l}-N_{r}\right)\xi_{0}v_{g}-\left(N_{l}+N_{r}\right)\xi_{0}\dot{L}.
\end{equation}
which $\dot{L}$ is the spindle's velocity.
Beside the above force, the effects due to the complex properties of cellular fluid should be added. The viscoelastic properties of the cytoplasmic fluid can be modeled by defining an effective
viscous drag coefficient (\emph{i.e.} $\xi$) and an effective elastic modulus for the spindle body (\emph{i.e.} $\kappa$). The effects of other astral microtubules in the real three
dimensional configuration are also included in this elastic constant, $\kappa$. As a result, the governing dynamical equation for the spindle body reads:
\begin{equation}\label{spindle}
\xi \dot{L}+\kappa(L-L_{0}) = F_{pol}.
\end{equation}
The elastic term acts as a centering force, it enforces the spindle body to choose an equilibrium position $L_0$.
Due to the dynamic instability of microtubules structure,
 a growing microtubule may change to a shrinking one even if it
has reached the cortex surface. However, this happens with a new
catastrophe frequency, compared to its bulk value. Then, the number of attached microtubules at both sides obey the following equations:
\begin{eqnarray}\label{numbers on cortex}
\dot{N}_{l}=-f_{wall}N_{l}+(v_{g}-\dot{L})N n^{+}_{l}(L,t),\nonumber\\
\dot{N}_{r}=-f_{wall}N_{r}+(v_{g}+\dot{L})N n^{+}_{r}(L,t).
\end{eqnarray}
which $N$ is the total number of microtubules.
In the limiting case, where the spindle body doesn't move, these equations are the same as  Eq.~\ref{wallboundary} multiplied by  $N$,
as a total number of microtubules .

To solve above dynamical equations and consequently obtain the dynamical behavior of spindle body, we assume that the system eventually
reaches an oscillating steady state that is a steady state solution which oscillates around an equilibrium state. We denote the equilibrium value of spindle position by $L_0$ and the number of attached polymers by $N_{l}^{0}$ and $N_{r}^{0}$.
Assuming that the polymerization forces in comparison with the elastic restoring forces  are small, we are looking for
perturbative solutions for spindle body in the following form:
\begin{equation}
L=L_{0}+L_{1}e^{i\omega t},
\end{equation}
where $L_1$ is a small deviation from the equilibrium position. In this case the number of attached microtubules can be expanded as below:
\begin{equation}\label{oscillations}
N_{l}=N_{l}^{0}+N_{l}^{1}e^{i\omega t},~~~N_{r}=N_{r}^{0}+N_{r}^{1}e^{i\omega t}.
\end{equation}
We note that the time scale for the polymerization process is smaller enough than the time scale for the spindle body
motion. This fact allows us to assume that for a definite value of $L(t)$, the number of attached microtubules is related to the length $L(t)$ through the steady state relations presented in Eq.~\ref{equilibriumvalue}. We also expand the number density up to the first order of
perturbation, for zeroth order value, we  can use the time
independent steady value (\textit{i.e.} no oscillation). Now, to find number densities on cortex, we assume:
\begin{eqnarray}
n^{+}_{l}(L,t)=n^{+}_{l}(L_{0})+L_{1}e^{i\omega t}\frac{d}{dx}n^{+}_{l}(x)_{x=L_{0}},\\
n^{+}_{r}(L,t)=n^{+}_{r}(L_{0})-L_{1}e^{i\omega t}\frac{d}{dx}n^{+}_{f}(x)_{x=L_{0}},
\end{eqnarray}
then we will have:
\begin{eqnarray}
&&n^{+}_{l}(L,t)-n^{+}_{r}(L,t)=-2
\frac{L_{1}}{\lambda}n^{+}(L_{0})e^{i\omega t},
\end{eqnarray}
and
\begin{eqnarray}
&&n^{+}_{l}(L,t)+n^{+}_{r}(L,t)=2n^{+}(L_{0}).
\end{eqnarray}
If we put the above information in Eq.(\ref{Fpol}) and Eq.(\ref{numbers on cortex}), we may derive $F_{pol}$ which can be used in
Eq.(\ref{spindle}). In the first order of perturbation expansion, our governing dynamical equation will be:

\begin{equation}
\left(\chi_{0}-i\chi_{1}\omega+\chi_{2}\omega^2\right)L_{1}=0.
\end{equation}
As one can see the overall motion of the spindle body is determined by a viscose term $\chi_1$ and an inertial term $\chi_2$. These effective
parameters are given by:
\begin{eqnarray}
\chi_{0}&=& f_{wall}(v_{g}\lambda B-\frac{\kappa}{\xi}),~\chi_{1}=f_{wall}(1+2B)+\frac{\kappa}{\xi}, \\
\chi_{2}&=& 1+B,~~~B= \frac{2\xi_{0}v_{g}Nn^{+}(L_{0})}{\xi f_{wall}}.
\end{eqnarray}
To investigate the existence of any oscillating state with nonzero amplitude, we can solve the above equation and find the allowed values for
$\omega$. Separating the real and imaginary parts of the frequency as $\omega=\omega_r+i\omega_i$, we can study the motion for different values of parameters.

\begin{figure}
\includegraphics[width=.8\columnwidth]{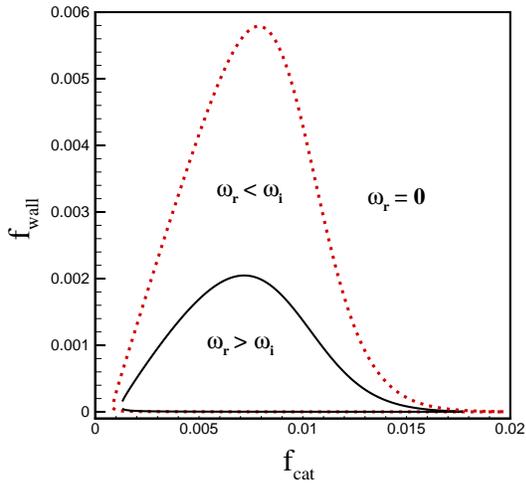}
\caption{(color online). The phase diagram for different values of $f_{wall}$ and $f_{cat}$ for a given experiment\cite{experiment}.
Three different states are separated. For states with $\omega_r=0$, the oscillating motion is not possible and all solutions are damped to zero,
while for $\omega_r\neq 0$, oscillating solutions are possible. For $\omega_r >\omega_i$, the damping time scale is large enough to
have a well defined oscillation.
}
\label{fig2}
\end{figure}

\begin{figure}
\includegraphics[width=.8\columnwidth]{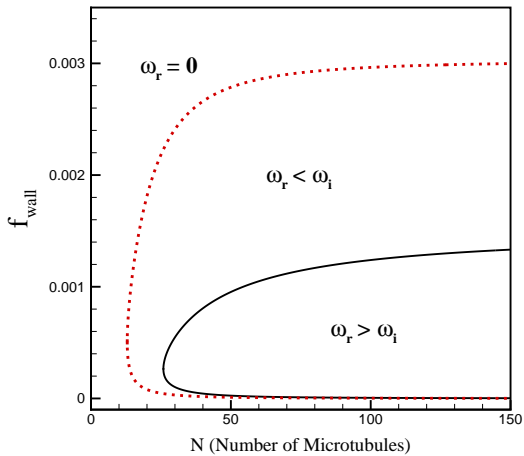}
\caption{(color online). The phase diagram for different values of total number of microtubules $N$ and $f_{wall}$ for a given experiment\cite{experiment}.
Three different states are separated. For states with $\omega_r=0$, the oscillating motion is not possible and all solutions are damped to zero,
while for $\omega_r\neq 0$, oscillating solutions are possible. For $\omega_r >\omega_i$, the damping time scale is large enough to
have a well defined oscillation.
}
\label{fig3}
\end{figure}

Fig.~\ref{fig2}, and Fig.~\ref{fig3} show two different phase diagrams of the system.
Parameter values which we have used are: $v_{g} = 1.41 \mu m / min$, $v_{s} = 30.7 \mu m / min$,
$f_{res} = 0.015 s^{-1}$, $f_{cat} = 0.0037 s^{-1}$, $\kappa = 4 \times 10^{-6}
N/m$\cite{julicher}, $\xi = 10^{-6}  N s/m$\cite{julicher}, $K_{B}
T = 4.1  pN. nm$, $L_{0}= 10  \mu m $, $\nu = 0.01 s^{-1}$, $l=
0.6  n m$ and $k_{on} = 100 s^{-1}$ \cite{experiment}.
The phase diagram according to parameters $f_{wall}$ and $f_{cat}$ (Fig.~\ref{fig2}), shows that for catastrophe frequency on the wall
($f_{wall}$), below a critical value,
spindle can oscillate.
Fig.~\ref{fig3} shows the phase diagram for different values of the number of microtubules ($N$)
and catastrophe frequency on the wall ($f_{wall}$) . It can be seen from this graph that there is a critical
value for $N$ (depending on the value of $f_{wall}$), below it the oscillatory state is not present.

As an example for the numerical value of the spindle oscillation frequency (Fig. \ref{fig2}, Fig. \ref{fig3}), and for numerical values: $N=100$, $f_{wall}=0.001 s^{-1}$ and $f_{cat}=0.007 s^{-1}$,
the oscillating frequency for the systems is: $\omega_{r}=0.0023 ~s^{-1}$. Changing the wall catastrophe rate to $f_{cat}=0.005 ~ s^{-1}$, we
see that the oscillation frequency decrease to $0.0018 ~s^{-1}$.
Clearly by changing the unknown parameters, we can tune the frequency to capture the exact oscillation frequency
observed in live systems~\cite{julicher}.

\section{Microtubules, Motor Proteins and spindle}
\begin{figure}
\includegraphics[width=.8\columnwidth]{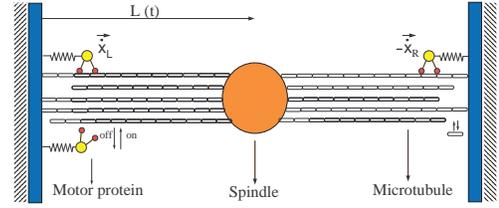}
\caption{(color online). One dimensional modeling of the  spindle, microtubules and motor proteins. Motor
proteins on the cell cortex can attach to the microtubules and exert force on spindle body. Microtubules polymerization also exerts force on the spindle and these two effects cause spindle to move.} \label{fig4}
\end{figure}

In this part we will consider the effect of motor proteins
\cite{julicher} as well as microtubules on the spindle's motion. As shown in
Fig.~\ref{fig4}, microtubules grow from the centrosome toward the wall and
 when they reach the wall, they exert force on it. Motor
proteins on the cortex can bind to these microtubules and
exert force on them as well. We assume that some fraction of motor proteins on the cortex, can attach to those microtubules which are not farther away than a critical distance from the cortex.
In this model we consider both forces that are exerted from microtubules
and motor proteins and try to find the overall equation of
motion for the spindle body.

We first present a simplified and mean field version of the motor protein mechanism introduced by S. Grill {\it et al.} \cite{julicher}. A motor protein can attach to a filament, walks along it and exerts force on it.
Denoting the walking velocity by $v$ and corresponding force by $f$ and for small velocities, the linear response: $v=v_{0}\left(1-f/f_{0}\right)$ holds during this motion~\cite{julicher,howard}. Here $f_{0}$ is the
stall force by which the motor protein stops and $v_{0}$ is the
velocity of motor in the absence of any force. Denoting the elastic properties of motor protein with a single parameter $k$ and its extension by $x$, we can
write the force motor feels as $f=kx$ (Fig.\ref{fig4}).
Then the motors velocities on the right wall and left one become:
\begin{eqnarray}\label{force velocity relation for motor}
\dot{x}_{L}=v_{0}-\mu x_{L}+\dot{L},\\
\dot{x}_{R}=v_{0}-\mu x_{R}-\dot{L},
\end{eqnarray}
where $\dot{L}$ is the spindle velocity,
$x_{L}$ is the extension of the linker (spring which connects the motor protein to the cortex) on the left side, $x_{R}$ is
the extension of the linker on the right side and $\mu$ is equal to
$kv_{0}/f_{0}$. The motor protein can attach to microtubule with a
rate $\omega_{on}$ and detach from it with a rate
$\omega_{off}$. According to these rates, we can find the
fraction of bond motor proteins as:
\begin{eqnarray}\label{fraction of bound motor}
\dot{Q}_{bL}= -\omega_{off}Q_{bL}+\omega_{on} (1-Q_{bL}),\\
\dot{Q}_{bR}= -\omega_{off}Q_{bR}+\omega_{on} (1-Q_{bR}),
\end{eqnarray}
$Q_{bL}$ is the fraction of bound motors on the left side while
$Q_{bR}$ is the fraction of bound motors on the right
side. For simplicity we have ignored the diffusion of motor proteins on the filament. We can replace all
time scales corresponding to the walking processes on the filament by $\omega^{-1}_{off}$, the average time that a motor spends in
attached state.
Since the detachment rate ({\emph i.e.} $\omega_{off}$) depends on the force motor feels,
it depends on the linker's length. However the attachment rate ({\emph i.e.}
$\omega_{on}$) is controlled by the barrier that motor feels to connect to the filament and
local temperature, and hardly depends on linker's length\cite{julicher}. The force dependent detachment
rate is $\omega_{off}=\omega_{0} \exp(ka|x|/k_{B}T)$ where $\omega_{0}$ is the detachment rate for motor
proteins in the absence of any force, $k$ is the spring's constant, $x$ is the linker's length and $a$ is
the molecular length scale.\cite{julicher}

\begin{figure}
\includegraphics[width=.8\columnwidth]{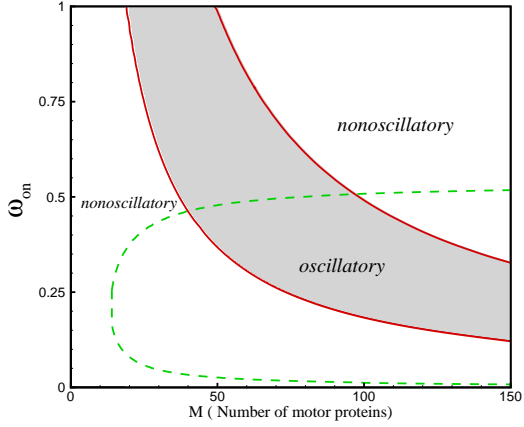}
\caption{(color online). Phase diagram of possible states for different values of the rate $\omega_{on}$ (in unit of $s^{-1}$),
and motor proteins number $M$. The colored region bounded by solid lines, shows the states of stable oscillatory
solutions for the case where the effects of  motor proteins and polymerization forces are included. Dashed line shows the
boundary for oscillatory and non oscillatory states when the polymerization force is not included \cite{julicher}.
Parameter values used in this graph are:
$k=8.3\times10^{-3} N/m$, $f_{0}=3 pN$,
$v_{0}=1.8 \mu m/s$
, $\omega_{0}=5 s^{-1}$\cite{julicher}, $L_{0}=10 \mu m$, $N=100$, $f_{wall}=0.001 s^{-1}$,
$v_{g} = 1.41 \mu m / min$, $v_{s} = 30.7 \mu m / min$, $f_{res} =
0.015 s^{-1}$ and $f_{cat}=0.0037 s^{-1}$\cite{dogterom1}.} \label{fig5}
\end{figure}
\begin{figure}
\includegraphics[width=0.8\columnwidth]{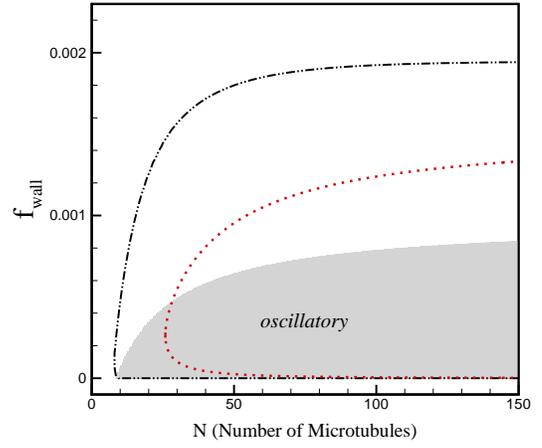}
\caption{(color online).
In this phase diagram we compare the results of a model that is based on the polymerization forces and a model that takes into account
both the polymerization forces and the effects due to the motor proteins.
As in Fig.~\ref{fig3}, the dotted line shows the boundary of different phases for the
case where only the polymerization effects are considered. Including the effects due to the motor proteins, we see that the dotted line changes to dashes-dotted-dotted line and also a new phase of
stable oscillating state ({\emph i.e.} $\omega_i<0$) comes to existence. This new sate is shown as a colored region.
For the physical parameters we
have used the following values:
$v_{g} = 1.41 \mu m / min$, $v_{s} = 30.7 \mu m / min$, $f_{res} =
0.015 s^{-1}$,$f_{cat}=0.0037 s^{-1}$\cite{dogterom1}, $k=8.3\times10^{-3} N/m$, $f_{0}=3 pN$,
$v_{0}=1.8 \mu m/s$
, $\omega_{0}=5 s^{-1}$, $\omega_{on}=0.25 s^{-1}$ \cite{julicher}, $L_{0}=10 \mu m$ and $M=100$.}\label{fig6}
\end{figure}

Two crucial elements in Grill \emph{et al.} mechanism are the stochastic nature of polymerization process and the
force dependent detachment rate that result oscillatory motion for the spindle. Keeping these two elements, we also have considered the force due to the microtubules polymerization. If microtubules reach a critical distance from the cortex, motor
proteins have the chance to attach to them with the rate $\omega_{on}$. If we denote by $P_{L}$ and $P_{R}$, the number of such microtubules
on the left and right sides, we can write:
\begin{eqnarray}\label{availableMT}
P_{L}=\int_{L-l_{0}}^{L}\left(n^{+}_{L}(x,t)+n^{-}_{L}(x,t)\right)dx,\\
P_{R}=\int_{L-l_{0}}^{L}\left(n^{+}_{R}(x,t)+n^{-}_{R}(x,t)\right)dx,\\
\end{eqnarray}
where $l_{0}$ is the critical distance from the cortex for a microtubule by which motor
proteins can attach to it. Here $n^{+}_{L}(x,t)$ ($n^{+}_{R}(x,t)$) is the number density of
growing microtubules on the left (right) side and
$n^{-}_{L}(x,t)$($n^{-}_{R}(x,t)$) is also the number density of
shrinking microtubules on the left (right)
side.

Combining the effects due to the polymerization and also the motor proteins, we can arrive
at the following dynamical equation for the spindle body:
\begin{eqnarray}\label{spindle motion}
\xi \dot{L}+\kappa(L-L_{0}) =
(N_{L}-N_{R})\xi_{0}v_{g}-(N_{L}+N_{R})\xi_{0}\dot{L}\nonumber\\
\qquad+M_{R}P_{R}Q_{bR}kx_{R}-M_{L}P_{L}Q_{bL}kx_{L},
\end{eqnarray}
where $N_{L}$ and $N_{R}$ are  the total number of microtubules in the left and right sides and $M_{L}$ and $M_{R}$ are
the total number of motor proteins in the left and right sides.

Now we can solve Eq.~\ref{numbers on cortex} and
Eq.~\ref{force velocity relation for motor}-\ref{spindle
motion}, and obtain the overall dynamical properties of the spindle body. Following the same perturbative method described in the previous section, we can arrive at the following equation in Fourier space:
\begin{equation}
\chi(\omega)\times L_1=0,
\end{equation}
where the dispersion $\chi(\omega)$ is given by:
\begin{equation}
\chi(\omega)=\chi_{4}\omega^4 + \chi_{3} \omega^3+ \chi_{2} \omega^2 + \chi_{1}\omega + \chi_{0},
\end{equation}
and the coefficients are given by:
\begin{eqnarray}
\chi_{4}&=&1+B,~\chi_{0}=\mu C (E f_{wall}-v_{g} \lambda B),\nonumber\\
\chi_{3}&=& (1+B)(H+f_{wall})+B+E+F,\nonumber\\
\chi_{2}&=&(1+B)(\mu C+f_{wall}H)+E(H+f_{wall})\nonumber\\
&&-v_{g}\lambda B+BH+F(f_{wall}+C)-G,\nonumber\\
\chi_{1}&=&(1+B)\mu C f_{wall} +E(\mu C+Hf_{wall})\nonumber\\
&&-v_{g}\lambda B H+\mu C B+FBf_{wall}-Gf_{wall},\nonumber\\
B&=& \frac{2\xi_{0}v_{g}Nn^{+}(L_{0})}{\xi f_{wall}} ,~~ C=\omega_{on} +\omega_{off}(x_{0}),\nonumber\\
D&=& \frac{\omega_{on}}{C} ,~~ E=2MD \frac{kv_{0}}{\mu \xi} \lambda P_{0} + \frac{\kappa}{\xi},\nonumber\\
F&=&\frac{2k M P_{0} C}{\xi} ,~  G=\frac{k v_{0}}{\mu} \omega^{'} (x_{0})F,\nonumber\\
H&=&\mu+C, ~ \mu= \frac{k v_{0}}{f_{0}}.\nonumber
\end{eqnarray}
To analyze the possible states we investigate the numerical solutions to $\chi(\omega_r+i\omega_i)=0$.
Fig.~\ref{fig5}, shows the phase diagram of different states of the spindle for different values of $M$, number of motor
proteins, and $\omega_{on}$, attachment rate for motor proteins. There
is a range of values for $\omega_{on}$ and $M$ that spindle has stable oscillations. This graph corroborates our main idea that microtubules polymerization is not negligible and changes the quantitative behavior of oscillation.

In Fig.~\ref{fig6}, we have shown the phase diagram of the system for different values of $f_{wall}$,
catastrophe frequency on the wall, and $N$, number of microtubules.
In this graph, we compare the results of a model that is based on the polymerization forces and a model that takes into account
both the polymerization forces and the effects due to the motor proteins.
As in Fig.~\ref{fig3}, the dotted line shows the boundary of different phases for the
case where only the polymerization effects are present. Including the effects due to the motor proteins, we see that
the dotted line changes to the dashed-dotted-dotted line and a new phase of
stable oscillating states(colored region) comes to existence. As an example of the oscillating frequency for $f_{wall}=0.0006 ~s^{-1}$, $M=100$, $N=100$ and  $\omega_{on}=0.25 ~s^{-1}$
we find that $\omega_r=0.006 ~ s^{-1}$ and $\omega_i=-0.0005 ~s^{-1}$.

\section{conclusion}
Similar to the Grill \emph{et al.}'s mechanism, we believe that the stochastic nature of the polymerization process and also the force dependent polymerization velocity are two important points that derive the oscillations in our model.
The effects due to the motor proteins are also considered in the frame work of a simplified version of the mechanism presented in \cite{julicher}. We have added both the polymerization forces and motor forces to obtain a more realistic description of the spindle motion. Taking into account these two mechanisms, we have studied the phase diagram for spindle dynamics.
 We have presented detailed phase diagrams for spindle motion and compared the phase diagram with and without the microtubules effect (Fig.\ref{fig5}). We see that adding microtubules polymerization effect changes the phase diagram; the critical number of motor proteins for which oscillation starts, are different in two cases. There is also a critical number for microtubules, above which the oscillation occurs. This critical amount changes if we consider/ignore effect of motor proteins. The numerical values for the oscillation frequency are in the range of observed experimental values.

In conclusion, while our results confirm Grill {\em et.al.}\cite{julicher} former idea that motor proteins are the main source of spindle oscillation, we see that microtubules polymerization has non-negligible effect on the quantitative behavior of such oscillation and can not be ignored.

S.R.S thanks A. Naji, S.N. Rasuli, L. Mollazadeh and P. Sens., A.N. thanks F. Julicher for stimulating discussions. A.N. acknowledge financial support from MPIPKS.

\end{document}